\documentclass[11pt]{article}

\usepackage[margin=2.5cm,top=1.5cm,bottom=1.5cm]{geometry}

\usepackage{graphicx}
\usepackage{blindtext}
\usepackage[textwidth=8em,textsize=small]{todonotes}
\usepackage{amsmath,multirow,float}
\usepackage{natbib,anyfontsize}
\usepackage{booktabs,geometry}
\usepackage{tabularx,array}
\usepackage{makeidx,url,algorithmicx,algorithm,algpseudocode,bm}
\usepackage{amssymb,color,bbm,bm}  

\newcommand{\one}[1]{\pmb{1}_{#1}}

\newcommand{\sm}{\overset{iid}{\sim}}

\def\x{\mathbf{x}}
\def\X{\mathbf{X}}

\def\R{\mathbb{R}}

\def\I{\mathrm{I}}
\def\W{\mathrm{W}}
\def\G{\mathrm{G}}
\def\N{\mathrm{N}}

\title{Bayesian forecasting of mortality rates using latent Gaussian models}

\author
{
    A. Alexopoulos
    \thanks{MRC Biostatistics Unit,
        University of Cambridge, UK.
                Email: {\tt angelos@mrc-bsu.cam.ac.uk}.
        }
    \and    
         P. Dellaportas
    \thanks{Department of Statistical Science, University College London, UK.
                Email: {\tt p.dellaportas@ucl.ac.uk}.
        } 
  \and           
       J.J. Forster
   \thanks{Department of Mathematical Sciences, University of Southampton,UK. 
               Email: {\tt J.J.Forster@soton.ac.uk}.
        }
   }

\begin{document}
\maketitle

\begin{abstract}
We provide forecasts for mortality rates by using two different approaches. First we employ dynamic non-linear logistic models based on Heligman-Pollard formula. Second, we assume that the dynamics of the mortality rates can be modelled through a Gaussian Markov random field. We use efficient Bayesian methods to estimate the parameters and the latent states of the proposed models. Both methodologies are tested with past data and are used to forecast mortality rates both for large (UK and Wales) and small (New Zealand) populations up to $21$ years ahead. We demonstrate that predictions for individual survivor functions and other posterior summaries of demographic and actuarial interest are readily obtained. Our results are compared with other competing forecasting methods.
\end{abstract}

\section{Introduction}

\subsection{Problem Setting}
Analysis of mortality data has long been of interest to actuaries, demographers and statisticians.  The first life tables were developed in the $17$th century, see for example \citet{graunt1977natural}.
What is perhaps the best-known mortality function is the analytical formula suggested by Benjamin Gompertz in 1825 \citep{smith1977mathematical}, which in many cases gives surprisingly good fits to empirical adult mortality rates. The earliest attempt to represent mortality at all ages is that of \citet{thiele1871mathematical}, who combined three different functions to represent death rates among children, young to middle-aged
adults, and the elderly, respectively. They proposed negative and
positive exponential curves for the first and third components and a
normal curve for the second.

Over a century later,  \citet{heligman1980age} used
a similar mathematical function that appears to provide
satisfactory representations of a wide variety of mortality patterns
across the entire age range. 

\begin{figure}
\centering
\includegraphics[scale=0.4]{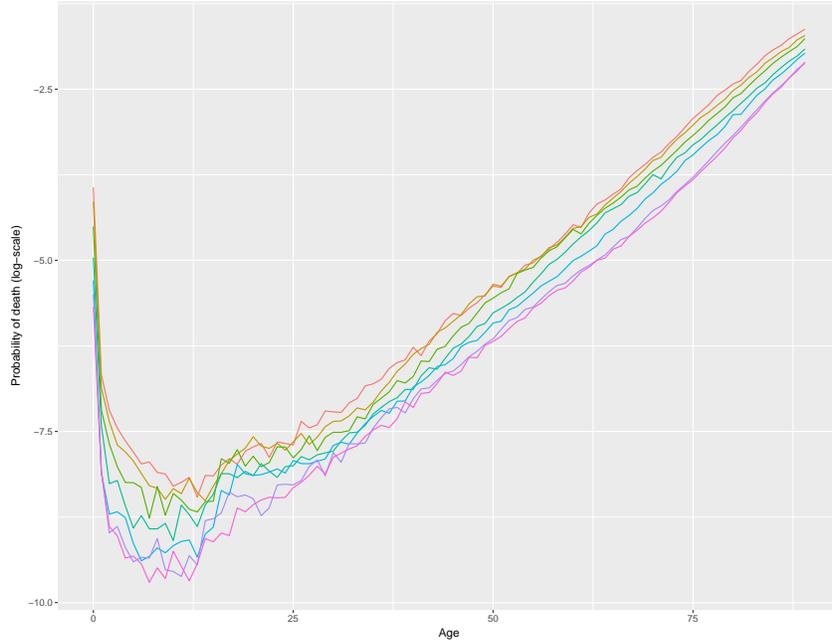}
   \caption{From top to bottom: log-probabilities of death versus age, for the years $1960$, $1970$, $1980$, $1990$, $2000$, $2010$ and $2013$ for females from UK-Wales.}
   \label{probdesc}
\end{figure}

Demographers, economists and social scientists are interested not only on the actual demographic structure of a country, but also on projections into the future. Although the static problem is rather straightforward, obtained readily from consensus data, the dynamic problem is a challenging problem with only partially satisfactory solutions.
A wide variety of mortality projection models are now available for practitioners, see
for example \citet{lee1992modeling}, \citet{brouhns2002poisson}, \citet{currie2004smoothing}, \citet{renshaw2006cohort}, \citet{cairns2006two}, and \citet{delwarde2007smoothing}.
The approach adopted until now is to select a single model, based on considerations of goodness-of-fit, past practice or other considerations, and project forward in time to produce not only expected future mortality 
rates but also an estimate of the associated uncertainty in the form of a prediction interval. For a visual illustration of the problem consider the mortality data of UK-Wales, obtained from the \cite{HMD}, between the years $1960-2013$ depicted in Figure \ref{probdesc}. Clearly the death probabilities are decreasing over the years and it is of particular interest to predict future mortality curves.

In what follows, $m_{zt}$ is used to represent the average, over time $t$, of the instantaneous death rate amongst the individuals with age in the interval $[z,z+1)$;  with $n_{zt}$ and $d_{zt}$ we denote the population at risk and the number of people who die at time $t$ with age in the interval $[z,z+1)$; and following \cite{currie2016fitting} we define the mortality rate $p_{zt}$ to be the probability of dying within one year for a person aged $z$ at time $t$.
The density of a $u$-variate Gaussian random 
variable $\X=(X_1,\ldots,X_u)$ with mean $\bm{\mu}$ and covariance matrix $\bm{S}$ evaluated at $\X$ is denoted by  $\phi_u(\X;\bm{\mu},\bm{S})$. Furthermore $\phi_u(\X;\bm{\mu},\bm{S},\bm{\eta},\bm{\xi})$, where $\bm{\eta} = (\eta_1,\ldots,\eta_u)$ and $\bm{\xi}=(\xi_1,\ldots,\xi_u)$, denotes the density of $\X$ conditional on the event that $X_i \in [\eta_i,\xi_i]$, $i=1,\ldots,u$ and $\eta_i, \xi_i$ are either real numbers or $-\infty$, $+\infty$ respectively; $\N_u(\bm{\mu},\bm{S} ; \bm{\eta}, \bm{\xi})$ denotes the corresponding $u$-variate truncated Gaussian distribution.
By assuming that we have past data containing the number of people being at risk at time $t$ aged $z$ and the corresponding number of deaths $d_{zt}$, our interest lies on forecasting the values $p_{z(T+1)}, p_{z(T+2)},\ldots$.

\subsection{A review of modelling and forecasting mortality rates}
Useful review material and case studies 
comparing models are provided by \cite{booth2008mortality}, \cite{cairns2011mortality} and \cite{haberman2011comparative}. 
Here we categorize mortality models 
into three main types. 

\subsubsection{Lee-Carter model and extensions}
The best known mortality model, and most successful in terms of generating extensions is the Lee-Carter model \citep{lee1992modeling} which models the logarithm of $m_{zt}$ as a bilinear function of age and time, that is
\begin{equation}\label{eq:lee}
\log m_{zt} = a_z+\beta_z\zeta_t  
\end{equation}
where $a_z$, $\beta_z$ and $\zeta_t$ are parameters to be estimated from relevant data.
A time series model is used for $\zeta_t$, which allows projections to be made using estimates of future $\zeta_t$ based on the corresponding time series forecast. \cite{renshaw2003lee} add flexibility to the model by incorporating a second bilinear term on the right hand side of \eqref{eq:lee}.

The original Lee-Carter model fits parameters by least squares methodology based on observed log-death rates 
(implicitly assuming a lognormal model for observed death rates). More satisfying and justifiable 
statistically are approaches which use \eqref{eq:lee} 
as a component of a Poisson model (possibly allowing also for overdispersion) for the observed 
numbers of deaths, as originally 
suggested by \cite{brouhns2002poisson}.

Various extensions of the basic Lee-Carter model have been proposed, most notably 
the introduction of cohort effects, see \cite{renshaw2006cohort}, where \eqref{eq:lee} is modified to
\begin{equation}\label{eq:cohort}
\log m_{zt} = a_z+\beta_z^{(0)}\gamma_{t-z}+\beta_z^{(1)}\zeta_t  
\end{equation}
where $\beta_z^{(0)}\gamma_{t-z}$ represents a bilinear effect depending on cohort $(t-z)$.

The basic Lee-Carter model does not impose any smoothness on the age parameters $a_z$ and $\beta_z$, 
which particularly in the case of $\beta_z$ can result in estimates which are unrealistic 
as functions of $z$. Approaches to overcome this problem involve smoothing the age parameters, either explicitly by constructing a smooth parametric model \citep{de2006extending} or by imposing a priori smoothing constraints on the parameters either via penalized maximum likelihood estimation \citep{delwarde2007smoothing} or, in a Bayesian framework, via an hierarchical prior distribution \citep{girosi2008demographic}. A related approach proposed by \cite{hyndman2007robust} smooths the observed $\log m_{zt}$ data using standard non-parametric smoothing techniques and then fits a functional regression model to the smoothed data using a set of orthonormal basis functions of age. The corresponding functional regression coefficients are time-varying and projected using a time series model.  
Recently, \cite{li2013extending} proposed also some extensions to the basic Lee-Carter model. First, following \cite{li2005coherent}, they modified the Lee-Carter method in order to produce projections that are non-divergent between the two sexes. Then, they extended the model to account for changes in the age specific rates of mortality-decline over the years. They model the fact that mortality-decline is decelerating at younger ages and accelerating at old ages \citep{bongaarts2005long} by modelling $\beta_z$ to depend on time $t$ through suitable functions.  They note that their model is particularly useful for projections over very long time horizons, while it reduces to the Lee-Carter method for less than $80$ years ahead predictions.

\subsubsection{Generalized linear models}
Several approaches have been proposed in which the bilinear term in \eqref{eq:cohort} is replaced by linear terms, the simplest of these being the classic age-period-cohort (APC) model 
\begin{equation}\label{eq:apc}
\log m_{zt} = a_z+\beta_t+\gamma_{t-z}  
\end{equation} 
which is commonly used in demographic and epidemiological applications.

\cite{renshaw2003lee} proposed (variations of) a model which can be expressed as
\begin{equation*}
\log m_{zt} = a_z+\beta_zt+\gamma_{t}  
\end{equation*} 
where the $\gamma_t$ are used in modelling observed data, but implicitly set to zero for future projections. \cite{cairns2006two} proposed the logistic-linear model \begin{equation}\label{eq:cairns}
\log \frac{p_{zt}}{1-p_{zt}} = \zeta_t^{(1)} + \zeta_t^{(2)}(z-\bar{z}) 
\end{equation} 
where $(\zeta_t^{(1)},\zeta_t^{(2)})$ are modelled as a bivariate random walk. Extensions to this model are presented and compared by \cite{plat2009stochastic}, \cite{cairns2011mortality} and \cite{haberman2011comparative}.

A generalized linear model which is not directly based on the Lee-Carter formulation is proposed by \cite{currie2004smoothing} and extended by \cite{kirkby2010smooth}. Here $\log m_{zt}$ is modelled as a smooth function in two dimensions (age and time) by using a generalized linear model with covariates derived from a (product) spline basis. Estimation is performed by penalized maximum likelihood, the penalty function imposing smoothness by penalizing discrepancies between neighbouring spline coefficients. 

\subsubsection{Non-linear models}
Various models have been proposed where mortality is expressed as a parametric function of age. Perhaps the best known of these is the Heligman-Pollard model \citep{heligman1980age} where the odds of death as a function of age is
\begin{equation}\label{eq:HelPol}
\frac{p_z}{1-p_z}=A^{(z+B)^{C}}+De^{-E(\log(z)-\log(F))^2}+GH^z
\end{equation}
where $A,B,C,D,E,F,G,H$ are unknown parameters.
Parameters $A,B,C,D$ take values in the interval $(0,1)$, while for the parameters $E$ and $F$ we have that $E \in (0,\infty)$ and $F \in (10,40)$.
Finally, $G \in (0,1)$ and $H \in (0,\infty)$, see \cite{dellaportas2001bayesian} for a more detailed discussion.
\cite{rogers1986parameterized} and
\cite{congdon1993statistical} have noted that estimation of the parameters of the Heligman-Pollard model is problematic because of the overparameterization of the model.  \cite{dellaportas2001bayesian} discuss the use of weighted least squares for the estimation of the Heligman-Pollard model and suggest Bayesian inference through a Markov Chain Monte Carlo (MCMC) algorithm.
Forecasting the future is more involved.
The approach adopted until now is to first estimate the parameters of the model for each age and for each year interval and
then to model the estimated parameters via a time series model. 
Clearly, such approaches ignore the parameter uncertainty as well as the parameter dependence. These approaches have been adopted by \cite{forfar1985changing}, \cite{rogers1986parameterized}, \cite{mcnown1989forecasting}, \cite{thompson1989multivariate} and \cite{denuit2009life}.

\cite{sherris2011modeling} describe an approach to mortality forecasting by fitting a Heligman-Pollard model to the death probabilities $p_{zt}$, over time, with time varying parameters $A_{t}$, $B_{t}$, $C_{t}$, $D_{t}$, $E_{t}$, $F_{t}$, $G_{t}$, $H_{t}$. A vector autoregression is used to model and project these time-varying estimated parameters in order to obtain mortality projections.

\subsection{Our contribution}
We propose two modelling approaches to perform our predictions. First we generalize the work of \citet{dellaportas2001bayesian} by including a dynamic component in their model based on the Heligman-Pollard formula. We assume that the eight parameters of the model evolve as a random walk parameters, thus relaxing any stationarity assumptions for the characteristics of the mortality curve. Second, we propose the use of a non-isotropic Gaussian Markov random field (GMRF) on a lattice constructed with ages $z$ and years $t$ and we project to the future by exploiting the estimated past features of the process. For both of the proposed models we use Bayesian methods to estimate their latent states and their parameters. More precisely, both models belong to the class of latent Gaussian models. The models consist of a non-normal likelihood and a Gaussian prior for their latent states. Bayesian inference for this type of models relies on an MCMC algorithm which alternates sampling from the full conditional distributions of the parameters of the model and the vector of the latent states.  

The step of sampling from 
the full conditional distribution 
of the parameters is usually conducted either directly or by using simple Metropolis-Hastings (MH) updates. 
The step of sampling the latent 
states of the model is challenging, since it 
usually consists of sampling from a 
distribution which is high 
dimensional and non-linear, 
see for example 
\cite{carter1994gibbs}, 
\cite{gamerman1997sampling}, \cite{gamerman1998markov}, 
\cite{knorr1999conditional} and 
\cite{knorr2002block} for some earlier attempts for Bayesian inference for 
the latent states of latent Gaussian models. 
However it is recognised  \citep{cotter2013mcmc} that a MH step targeting the conditional distribution of the latent states of a latent Gaussian model has to be both likelihood and prior informed. Proposals 
that are informed by the likelihood of a 
latent Gaussian model are proposals which are based on the discretization of the Langevin diffusion 
and they are used in the Metropolis adjusted 
Langevin algorithm (MALA) developed by 
\cite{roberts1996exponential} 
and the manifold MALA and Riemann 
manifold Hamiltonian Monte Carlo 
developed by \cite{girolami2011riemann}. 
Proposals that are taking 
into account the dependence structure of the Gaussian prior of the latent states have been 
designed by \cite{bernardo1998regression} 
and by \cite{murray2010slice}, 
see also \cite{beskos2008mcmc} for a detailed discussion. Finally, \cite{cotter2013mcmc} and \cite{titsias2018auxiliary} construct proposal distributions which are informed both from the likelihood and the prior. In this paper we construct proposals that exhibit these properties in both of the proposed models.

\subsection{Structure of the paper}
The paper is organized as follows. In Section 2 we present our model based on the Heligman-Pollard formula. In Section 3 we adopt our second approach in the problem where we use a non-parametric model based on Gaussian processes. In Section 4 we present the application of our models on the UK-Wales and New Zealand data and we compare it with other competing models. Section 5 concludes with a brief discussion.

\section{A dynamic model based on Heligman-Pollard formula}
\label{HPmodel}
In their paper, \citet{heligman1980age}, argue that a mortality
graduation can only be considered successful if the graduated rates progress smoothly from age to age and at the same time they reflect
accurately the underlying mortality pattern.  For this reason they
propose a mathematical expression or law of mortality which they
fit to post-war Australian national mortality data.

The curve that they suggest is given by equation \eqref{eq:HelPol}.
To define the dynamic version of the model, let $\bm{\psi}_t =(\tilde{A}_{t},\tilde{B}_t,\tilde{C}_{t},\tilde{D}_t,\tilde{E}_t,\tilde{F}_t,\tilde{G}_t,\tilde{H}_t)' $ be the latent states of the model parameters at time $t$, where the elements of $\bm{\psi}_t$ are obtained from the original variables using a suitable transformation so that $\bm{\psi}_t \in \R^8$. For example we set $\tilde{A}_{t} = \log(A_t/(1-A_t))$ and $\tilde{E}_{t} = \log(E_t)$. 
Throughout this paper, $t$ will refer to a year while $T$ is the number of years in the past for which we have data.
The odds of death at time point $t$ are assumed to be given by the Heligman-Pollard model:
\begin{equation}\label{eq:dynHP}
\frac{p_{zt}}{1-p_{zt}}=A_{t}^{(z+B_{t})^{C_{t}}}+D_{t}e^{-E_{t}(\log(z)-\log(F_{t}))^2}+G_{t}H_{t}^z 
\end{equation}
where $z=0,1,\ldots,\omega$, $t=1,\ldots,T$ and $\omega$ is the age of the oldest people in the data.
We denote the right side of \eqref{eq:dynHP} with $K(z,\bm{\psi}_t)$ and we have that 
\begin{equation}\label{eq:prob}
p_{zt}=\frac{K(z,\bm{\psi}_t)}{1+K(z,\bm{\psi}_t)}
\end{equation}
while the likelihood of our model is 
\begin{equation}\label{eq:like}
\pi(\bm{d}|\bm{\psi})=\prod_{t=1}^T\prod^{\omega}_{z=0}\binom{n_{zt}}{d_{zt}}K(z,\bm{\psi}_t)^{d_{zt}}[1+K(z,\bm{\psi}_t)]^{-n_{zt}}
\end{equation}
with $\bm{d}$ denoting the vector with elements $d_{zt}$ for $z = 0,1,\ldots,\omega$ and  $t=1,\ldots,T$.

For the dynamic modelling of the latent states in $\bm{\psi}_t$ we assume a random walk structure and we have that 

\begin{equation}\label{eq:truncrw} 
\pi(\bm{\psi}_{t}|\bm{\psi}_{t-1},\bm{\mu},\bm{\Sigma},\bm{\eta},\bm{\xi}) = \phi_8(\bm{\psi}_{t};\bm{\psi}_{t-1} + \bm{\mu}, \bm{\Sigma}, \bm{\eta}, \bm{\xi}),\,\,\, t=2,\ldots,T,
\end{equation}
$\pi(\psi_{i1}) \propto 1$ if $\psi_{i1} \in [\eta_i,\xi_i]$ and $\pi(\psi_{i1}) = 0$ otherwise, where $\psi_{it}$ denotes the $i$th element of $\bm{\psi_t}$, $i=1,\ldots,8$.

The random walk process defined by equation \eqref{eq:truncrw} imposes a lot of prior structure for the parameters of the Heligman-Pollard model and relaxes any stationarity assumptions for their evolution across the years. Specification of the vectors $\bm{\eta}$ and $\bm{\xi}$ allows the representation of our prior beliefs about the range of the parameters of the model and restricts known problems such as overparameterization \citep{congdon1993statistical}, non-identifiability \citep{bhatta2013bayesian} and change in age patterns of mortality-decline \citep{li2013extending} across the years. In our applications we fix the elements of the vectors $\bm{\eta}$ and $\bm{\xi}$ based on prior beliefs, expressed as $1\%$ and $99\%$ percentiles, reported by \cite{dellaportas2001bayesian} , by setting  $\bm{\eta} = (-10.61,-10.61,-5.99, -11.29,$ $-25.33,-\infty,-17.5,-1.39)'$ and $\bm{\xi} = (-2.75,-0.2,2.2,$
$-3.48,4.09,2.64,-3.48,0.18)'$.

For the drift $\bm{\mu}$ of the random walk process we assume that $\pi(\bm{\mu})=\phi_8(\bm{\mu};0, \bm{M}^{-1})$, where $\bm{M}$ is a diagonal  $8\times8$ matrix with elements equal to $0.001$. For the variance-covariance matrix $\bm{\Sigma}$ we assume the following inverse Wishart $(\I \W)$ prior suggested by 
\cite{huang2013simple},
$$
\bm{\Sigma}|\bm{\alpha} \sim \I \W (\nu+8-1, 2\nu \bm{\Sigma}_{prior})
$$
where $\bm{\alpha}=(\alpha_1,\ldots,\alpha_8)$, $\bm{\Sigma}_{prior}$ is a diagonal matrix with elements $1/ \alpha_1$ $,\ldots,$ $1/ \alpha_8$ in the diagonal, $(\nu+8-1)$ are the degrees of freedom of the inverse Wishart distribution and for the parameters $\alpha_i$ we assume the following inverse gamma $(\I \G)$ prior distributions
$$
\alpha_{i} \sm \I \G (1/2, 1/ \ell^2)
$$
for all $i=1,\ldots,8$ while, following \cite{huang2013simple}, we set $\ell = 10^5$.
The above prior structure implies half-$t(\nu,\ell)$ prior distributions for the standard deviations $\sigma_i$ in the diagonal of $\bm{\Sigma}$ and by choosing $\nu=2$ we have uniform, $U(-1,1)$, prior distributions for the correlation of the latent states in $\bm{\psi}_t$; see \cite{gelman2006prior} and \cite{huang2013simple} for a detailed presentation of this prior distribution for the covariance matrix $\bm{\Sigma}$. 
Denoting by $\theta = (\bm{\Sigma},\bm{\mu},\bm{\alpha})$ the parameters of the model and by $\bm{\psi}=(\bm{\psi_1}',\ldots,\bm{\psi_T}')'$ the latent states of the model the posterior distribution of interest is

\begin{equation}\label{eq:postH}
\pi(\bm{\psi},\theta |\bm{d},\bm{\eta},\bm{\xi}) \propto \pi(\theta)\pi(\bm{d}|\bm{\psi})\prod_{t=2}^T\phi_8(\bm{\psi}_t;\bm{\psi}_{t-1}+\bm{\mu},\bm{\Sigma},\bm{\eta},\bm{\xi}).
\end{equation}

By noting that any of the conditional distributions for the elements of $\bm{\psi}$ depends on the vectors $\bm{\eta}$ and $\bm{\xi}$, we simplify our notation and we drop reference to them for the remaining of the Section.

Our aim is to predict the probabilities $p_{zt}$  at some future time points $t=T+1,T+2,\ldots$, for all $z=0,1,\ldots,\omega$. To compute, for example, the posterior predictive distribution of $p_{z,T+1}$, 
we first have to approximate
\begin{equation}\label{eq:HPpred}
\pi(\bm{\psi}_{T+1}|\bm{d})=\int \pi(\bm{\psi}_{T+1}|\bm{\psi}_{T},\theta)\pi(\bm{\psi},\theta |\bm{d})d\bm{\psi} d\theta
\end{equation}
and then to compute the predictive density of $p_{z,T+1}$ based on the equation \eqref{eq:prob}. 
The integral in \eqref{eq:HPpred} is usually approximated as follows \citep{geweke2010comparing}.
First we have to obtain $M$ samples from the distribution with density $\pi(\bm{\psi},\theta |\bm{d})$ and then for each sample 
$\bm{\psi}^m,\theta^m$ we draw $\bm{\psi}^m_{T+1}$ from the distribution with density $\phi_8(\bm{\psi}_{T+1};\bm{\psi}_{T}^m + \bm{\mu}^m, \bm{\Sigma}^m,\bm{\eta},\bm{\xi})$. The 
values $\{\bm{\psi}^m_{T+1}\}^M_{m=1}$ form, through equation \eqref{eq:prob}, a sample from the posterior predictive distribution of $p_{z,T+1}$. The same procedure can be used for every future time point $T + 2, T + 3,\ldots$.

It is clear from \eqref{eq:postH} that the proposed model is a latent Gaussian model with latent states $\bm{\psi}$ and hyperparameters $\theta$. 
To obtain samples from \eqref{eq:postH} we construct a Metropolis within Gibbs sampler which alternates sampling from $\pi(\bm{\psi}|\theta,\bm{d})$ and $\pi(\theta|\bm{\psi},\bm{d})$. Sampling from $\pi(\theta|\bm{\psi},\bm{d})$ can be conducted directly since the full conditional distributions of the hyperparameters $\bm{\Sigma},\bm{\mu}$ and $\bm{\alpha}$ are of known form.
Sampling from $\pi(\bm{\psi}|\theta,\bm{d})$ is performed by using $T$ MH steps to update each $\bm{\psi}_t$. In Section $5$ of the on-line supplementary material we derive the required full conditional distributions. 

An important feature of the MH steps that we use to sample from the distribution with density $\pi(\bm{\psi}|\theta,\bm{d})$ is the following.
We incorporate information from the likelihood of our model into the proposal distributions of the MH steps by following \cite{dellaportas2001bayesian}. We propose for each $t=1,\ldots,T$ new states for $\bm{\psi}_t$ from a Gaussian distribution with mean $\bm{m}_t$ and covariance matrix $c_t\bm{V}_t$. The vector $\bm{m}_t$ and the covariance matrix $\bm{V}_t$ are the maximum likelihood estimators and covariance (inverse Hessian) matrix derived by using a non-linear weighted least squares algorithm with weights $w_{zt} = 1/q_{zt}^2$, where $q_{zt}$ are the empirical mortality rates, for the age $z$ at time point $t$, as suggested by \cite{heligman1980age}. Finally, $c_t$ are pre-specified constants, which are tuned to achieve better convergence behaviour measured with respect to sampling efficiency (percentage of accepted proposed moves). After the initial iteration, the mean vector of the proposal density is updated with the current sampled parameter vector.

Thus, we construct a likelihood-informed proposal distribution which enables us to jointly update the eight parameters of the model. These characteristics of the proposed MCMC algorithm accelerate the convergence of the corresponding Markov chain by overcoming problems such as the strong posterior correlation of the parameters of the Heligman-Pollard model reported by \cite{dellaportas2001bayesian}. In Section \ref{RealDemog} we apply the present methodology to the UK-Wales and New Zealand data. We evaluate the mixing properties of the proposed MCMC algorithm using the effective sample size (ESS) of the samples drawn from the posterior distributions of interest. The ESS of $M$ samples drawn using an MCMC algorithm can be estimated as $s^2M/ \gamma_0$ where $s^2$ is the sample variance of the samples and $\gamma_0$ is an estimation of the spectral density of the Markov chain at zero. In the on-line supplementary material we compare the ESS of samples drawn from the posterior in \eqref{eq:postH} using our proposed MH steps with the ESS of samples drawn using simple random walk MH steps.

\section{A non-parametric model}
\label{IGMRFmortality}
A Markov random field is a joint distribution for the variables $(x_1,\ldots,x_n)$ which is determined by its full conditional distributions with densities $\pi(x_i|\x_{-i})$ where $\x_{-i}$ $= (x_1,\ldots,x_{i-1},$ $x_{i+1},\ldots,x_n)'$. In the case where the conditional distributions are Gaussian distributions the Markov random field is called Gaussian (GMRF), see \cite{GMRFbook}. There is a strong connection between GMRFs and conditional autoregressive models \citep{besag1974spatial}.
 
A special case of GMRFs that we will use to model mortality rates are the intrinsic GMRF models, in which the precision (inverse covariance) matrix of the joint (Gaussian) distribution of the variables $(x_1,\ldots,x_n)$ is a singular matrix, since it does not have full rank. 
In Section $1$ of the on-line supplementary material we present further details of GMRF models.

\subsection{Modelling mortality rates using an intrinsic GMRF model}
\label{IGMRFmortality}
To model mortality rates based on the model with likelihood given by equation \eqref{eq:like} we transform the probability $p_{zt}$ of death at age $z$ in the $t$th year in the variable $x_{zt} = \log(p_{zt}/(1-p_{zt}))$ for each $z=0,\ldots,\omega$ and $t=1,\ldots,T.$
Denote by $\x_t = (x_{0t},\ldots,x_{\omega t})'$ and let $\x = (\x_1',\ldots,\x_T')'$ be an $(\omega+1)T$-dimensional vector. It is useful to think a lattice with $(\omega+1) \times T$ nodes and $(z,t)$ denoting the element of the $z$th row and the $t$th column.
For the vector $\x$ we assume that it has a $(\omega+1)T$-variate Gaussian distribution with mean $\bm{\mu} = (b\one{\omega+1},$ $2b\one{\omega+1},\ldots,Tb\one{\omega+1})'$, where $\one{\omega+1}$ is a $(\omega+1)$-dimensional vector with ones, and precision matrix  
\begin{equation}\label{eq:matrix}
\bm{Q} = \tau(\rho_{age}\bm{R}_{\omega+1}\otimes \bm{I}_{T} + \rho_{year}\bm{I}_{\omega+1}  \otimes \bm{R}_{T})
\end{equation}
where $\bm{I}_{\omega+1}$ is the identity matrix of dimension $(\omega+1) \times (\omega+1)$ and $\bm{R}_{\omega+1}$
is $(\omega+1) \times (\omega+1)$ matrix  with elements
$R_{ij}$ defined as
 \begin{equation*}
R_{ij}=                                                               
\begin{cases}
1 & \text{if $i=1$ and $j=1$}\\
1 & \text{if $i=\omega+1$ and $j=\omega+1$}\\
2 & \text{if $i=j$ and $i,j \neq 1$ and $i,j \neq \omega+1$}\\
-1 & \text{if $\lvert i-j \rvert =1$}\\
0 & \text{otherwise}.
\end{cases}
\end{equation*} 
Following the described modelling perspective $\x$ is an intrinsic GMRF since $\bm{Q}$ is singular. It follows that for each $z=1,\ldots,\omega-1$ and $t=2,\ldots,T-1$, the full conditional density of $x_{zt}$ is normal with mean equal to
\begin{equation*}
\label{eq:condmean}
\frac{1}{4}(\rho_{age}(x_{z-1,t}+x_{z+1,t}) + \rho_{year}(x_{z,t-1}+x_{z,t+1}) )
\end{equation*}
and variance $1/4\tau$. The parameters $\rho_{age}$ and $\rho_{year}$ control the association of the death probabilities across ages and years respectively. We emphasize that $\rho_{age}$ and $\rho_{year}$ are expected to differ because they capture correlations across age and calendar time dimensions, while to guarantee model identifiability we assume that $\rho_{age} + \rho_{year} =2$.

\subsubsection{Bayesian inference}
\label{GPmodel}
The likelihood function of our model is given by the product of the terms in the right hand side of equation \eqref{eq:like}, 
\begin{equation}
\label{eq:GMRFlik}
\pi(\bm{d}|\x) = \prod_{t=1}^T\prod_{z=0}^{\omega} \binom{n_{zt}}{d_{zt}}p_{zt}^{d_{zt}}(1-p_{zt})^{n_{zt}-d_{zt}},
\end{equation}
where $\bm{d}$ is the $(\omega+1)T$-dimensional vector with elements $d_{zt}$ and $p_{zt} = \exp(x_{zt})/(1+\exp(x_{zt}))$.  By denoting by $\theta = (b, \rho_{age}, \tau)$ the parameters of the model we construct an MCMC algorithm that samples from the joint posterior distribution of the parameters and the latent states of the model which  has density 
$$
\pi(\theta,\x|\bm{d}) \propto \pi(\theta) \pi(\x|\theta)\pi(\bm{d}|\x),
$$
where $\pi(\theta)$ is the density of the prior distribution of the parameters, $\pi(\x|\theta)$ is the density of the (improper) $(\omega+1)T$-variate Gaussian distribution with mean $\mu$ and precision matrix $\bm{Q}$ and $\pi(\bm{d}|\x)$ is given by \eqref{eq:GMRFlik}.

Sampling from the distribution with density $\pi(\theta|\x,\bm{d})$ consists of sampling from the full conditional distributions of the parameters $b, \rho_{age}$ and $\tau$ of the model. In Section $4$ of the on-line supplementary material of this paper we present the densities of these full conditionals and we note that we can sample from these either directly $(\tau,b)$ or by random walk MH steps on $\rho_{age}$.

Sampling from $\pi(\x|\theta,\bm{d})$ consists of sampling from the distribution with density proportional to the product of the density of the $(\omega+1)T$-variate Gaussian prior of the latent states $\x$ and to the intractable likelihood given by \eqref{eq:GMRFlik}. We use the gradient-based auxiliary MCMC sampler proposed by \cite{titsias2018auxiliary} for sampling the latent states of the proposed model. In this case the gradient-based auxiliary sampler makes efficient use of the gradient information of the (intractable) likelihood and is invariant under the tractable Gaussian prior. \cite{titsias2018auxiliary} show, by conducting extensive experiments in the context of latent Gaussian models, that the gradient-based auxiliary sampler outperforms, in terms of the ESS, well established methods such as MALA \citep{roberts2002langevin}, elliptical slice sampling \citep{murray2010slice} and preconditioned Crank-Nicolson Langevin algorithms \citep{cotter2013mcmc}. Finally, an attractive feature of the gradient-based auxiliary sampler is that its implementation is straightforward and requires only a single tuning parameter to be specified, which can be estimated during the burn-in period.

The proposal developed by \cite{titsias2018auxiliary} is based on an idea first appeared in \cite{titsias2011riemann} and is constructed as follows. 
Auxiliary variables $\bm{u}_{\x} \in \R^{(\omega+1)T}$ are proposed from a Gaussian distribution 
\begin{equation*}
\label{eq:titsiasaux}
\N ( \x + (\delta/2)\nabla \log \pi(\bm{d}|\x,\theta), (\delta/2)\bm{I}_{(\omega+1)T}   )
\end{equation*}
where $\nabla \log \pi(\bm{d}|\x,\theta)$ denotes the gradient of the log-likelihood evaluated at the current states of $\x$ and $\theta$. Then new values $\x_{prop}$ are proposed from the distribution with density
\begin{equation}
\label{eq:titsiasaux2}
q(\x_{prop}|\bm{u}_{\x}) \propto \phi_{(\omega+1)T}(\x_{prop}; \bm{u}_{\x}, (\delta/2)\bm{I}_{(\omega+1)T})\pi(\x|\theta), 
\end{equation}
and the proposed value $\x_{prop}$ is accepted with MH acceptance probability $min(1, \alpha)$ given by
\begin{equation}
\label{eq:TratioIntro}
\alpha= \frac{\pi(\bm{d} | \x_{prop},\theta)}{\pi(\bm{d} | \x,\theta)}\exp\{  f(\bm{u}_{\x},\x_{prop}) - f(\bm{u}_{\x},\x)   \}
\end{equation}
and $f(\bm{u}_{\x},\x) = (\bm{u}_{\x} - \x - (\delta/4)\nabla \log \pi(\bm{d} | \x,\theta) )' \nabla \log \pi(\bm{d} | \x,\theta)  $, while \cite{titsias2011riemann} suggests to tune the parameter $\delta$ in order an acceptance rate of $50\%-60\%$ to be achieved.
In Section $2$ of the on-line supplementary material we summarize the steps of this algorithm.

For every $z$ and $k$, our aim is to predict the death probabilities $p_{zt}$  at future time points $t= T+1,T+2,\ldots,T+k$  expressed through  the vectors $\x^*=(\x_{T+1}',\ldots,\x_{T+k}')'$.  The required predictive density is 
\begin{equation}
\label{eq:GPpred}
\pi(\x^*|\bm{d} ) = \int \pi(\x^*|\x,\theta)\pi(\x,\theta|\bm{d})d\x d\theta.
\end{equation}
In Section $3$ of the on-line supplementary material of the paper we describe how we approximate the integral in \eqref{eq:GPpred} based on MCMC samples from the distribution with density $\pi(\x,\theta|d)$ and on properties of the multivariate normal distribution.  We evaluate this approximation by calculating the ESS of the drawn samples.  
This exercise confirms that our choice to use the MH proposed by \cite{titsias2018auxiliary} achieves Markov chains with good mixing expressed with high ESS. In the on-line supplementary material we present the ESS of the samples drawn from the posterior distribution of the latent states $\x$ of the model.

\section{Applications to real data}
\label{RealDemog}

\subsection{Prediction of mortality rates}
\label{Comp}

\begin{figure}
 \centering
\includegraphics[scale=0.7]{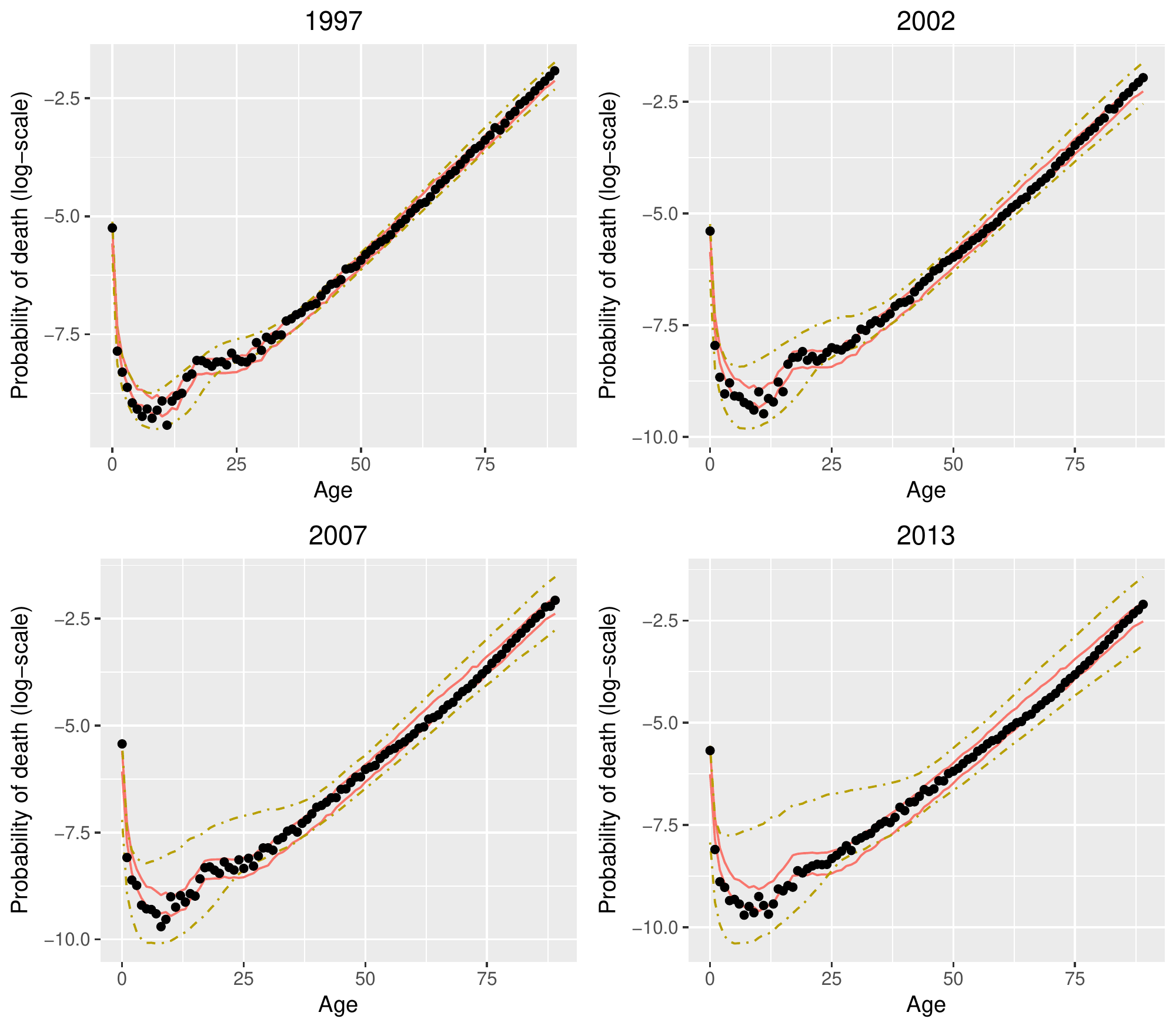}
	\caption{Predicted $95\%$  credible intervals for the  Gaussian Markov random field (red lines) and  the  Heligman-Pollard (brown dashed lines) models for UK-Wales mortality data based on observations for the years $1983-1992$. Black dots denote the true log-probabilities of death.  Top left: Predictions for 1997; Top right: Predictions for 2002; Bottom left: Predictions for 2007; Bottom right: Predictions for 2013.}
	\label{gphp}
\end{figure}

\begin{figure}
 \centering
\includegraphics[scale=0.7]{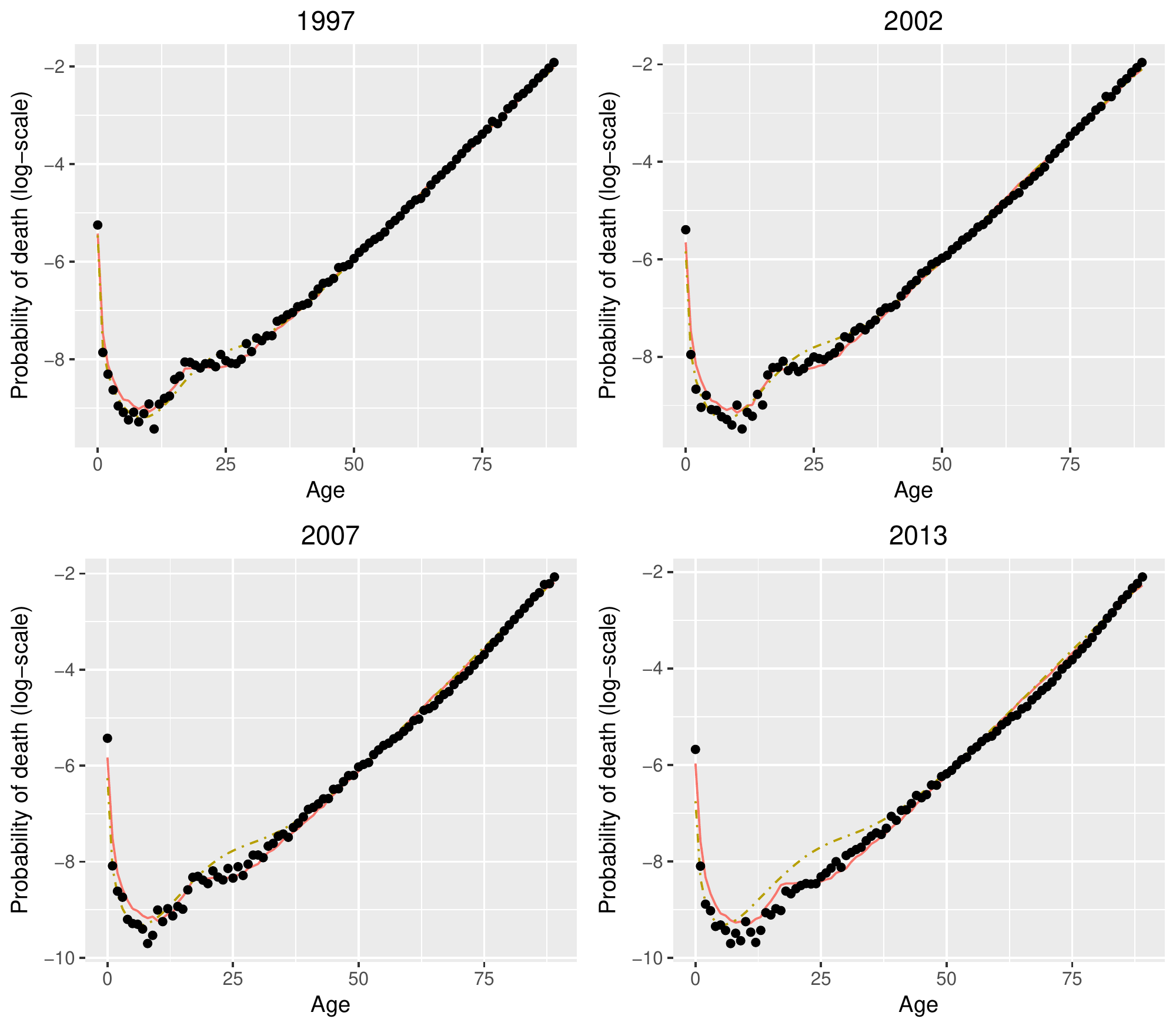}
	\caption{Predicted means for the  Gaussian Markov random
field (red lines) and  the  Heligman-Pollard (brown dashed lines) models for UK-Wales mortality data based on observations for the years $1983-1992$. Black dots denote the true log-probabilities of death.  Top left: Predictions for 1997; Top right: Predictions for 2002; Bottom left: Predictions for 2007; Bottom right: Predictions for 2013.}
	\label{PredMeans}
\end{figure}

Our suggested models express different modelling beliefs about the extrapolation of the mortality curve.  The Heligman-Pollard dynamic model suggests non-stationarity with variance increasing as the predictions move away in future, whereas the GMRF predictions are constrained by the strong Gaussian prior. To test how both models behave in real data, we  predict $5,10,15$ and $21$ years ahead mortality rates for UK-Wales based on observed data from \cite{HMD} during years $1983-1992$ $(T=10,\omega=89)$.  The results are compared with true observed mortality rates. 
Figure \ref{gphp} depicts  the  $95\%$ credible intervals of the posterior predictive distributions of the log-probabilities of death obtained from the Heligman-Pollard and the non-parametric models, while Figure \ref{PredMeans} presents the corresponding posterior means.  Both models perform well, with the Heligman-Pollard model achieving, as expected, wider credible intervals which are evaluated in Section \ref{comps} through a full-fledged quantitative evaluation. The proposed methods are not computationally expensive; our MCMC algorithms are written in R \citep{rcite} and we obtain $1,000$ iterations in $2.5$ minutes in the case of the GMRF model and in $6.5$ seconds in the case of the Heligman-Pollard model. Thus, we needed almost $8$ hours to complete $21$ years ahead predictions using the GMRF model and less than $4$ hours for the dynamic Heligman-Pollard model. However after fitting the two models in multiple datasets in Section \ref{comps}, we noted that the time for the Heligman-Pollard model varies between $4$ and $14$ hours depending on the dataset. See also in the on-line supplementary material where we provide details for the implementation of our algorithms.

\subsection{Prediction of survival probabilities}

\begin{figure}[t]
 \centering
\includegraphics[scale=0.5]{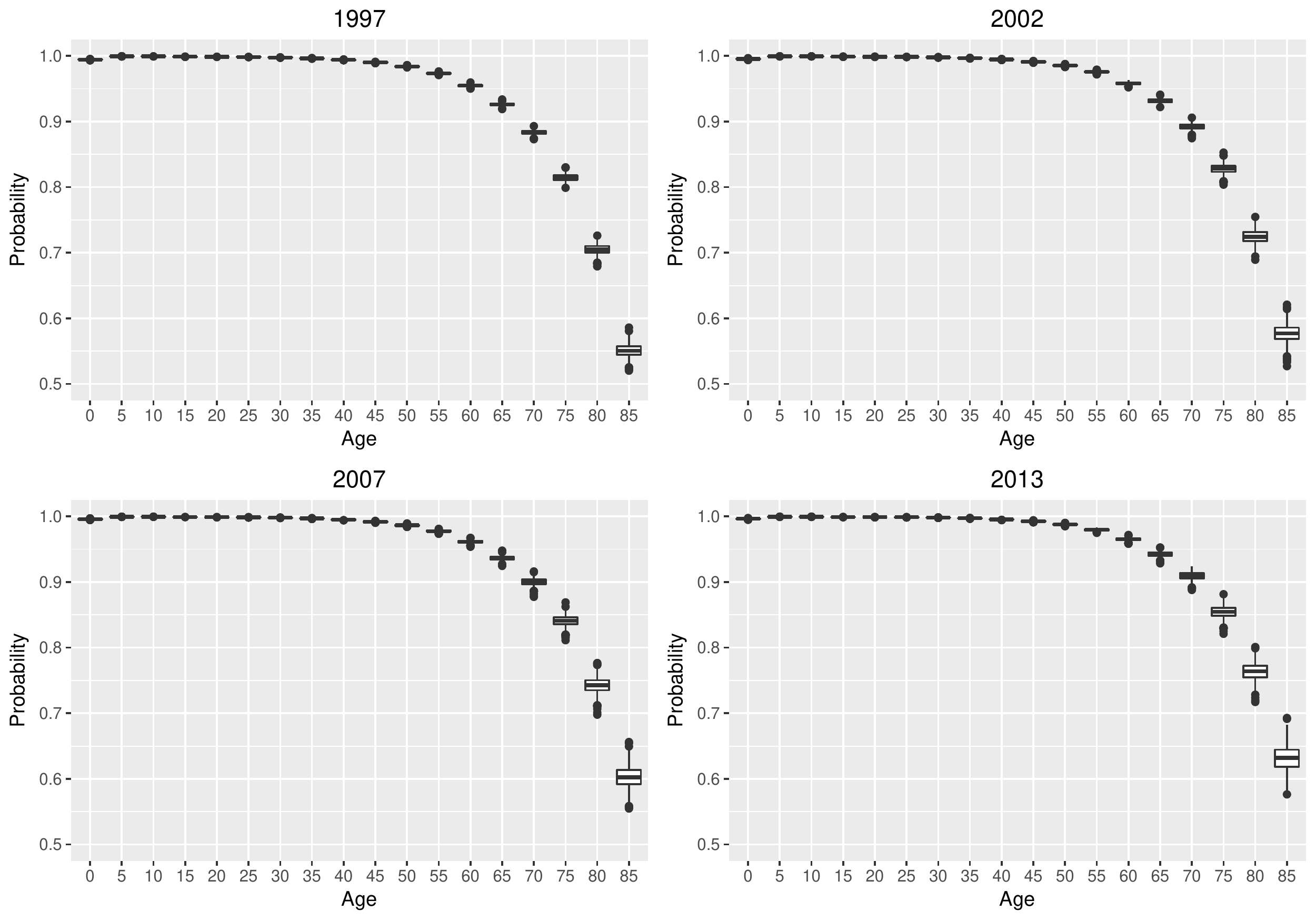}
	\caption{Posterior predictive distributions of survival probabilities ${}_{s}p_{z,T+k}$ for UK-Wales mortality data, with $s=5$, based on observations for the years $1983-1992$ ($T=10$) for ages $z=0,5,10,\ldots,85$ years old. Top left: Predictions for 1997 ($k=5$); Top right: Predictions for 2002 ($k=10$); Bottom left: Predictions for 2007 ($k=15$); Bottom right: Predictions for 2013 ($k=21$). The prediction of the corresponding death probabilities has been conducted using the GMRF model.}
	\label{surviveplot}
\end{figure}

An attractive feature of our Bayesian methods is that we can easily obtain prediction intervals for several quantities which are of interested to actuaries and demographers, but they are not readily available in non-Bayesian models.
Here we present projections of survival probabilities in an horizon of $k$ years ahead. These are defined as 
\begin{equation}\label{eq:survival}
{}_{s}p_{z,T+k} = \prod_{i=0}^{s-1}(1-p_{z+i,T+k})
\end{equation}
and denote the probability of a person aged $z$ at the year $T+k$ to survive up to age $z+s$. 
Following \cite{dellaportas2001bayesian} we utilize samples from the posterior predictive distributions of the probabilities of death in order to compute the probabilities in \eqref{eq:survival} for the data presented in Section \ref{Comp}. Figure \ref{surviveplot} summarizes the posterior samples of survival probabilities for $s=5$, projected in the years $1997,2002,2007$ and $2013$ ($k=5,10,15,21$) using the GMRF model. It is clear that we predict an increase of the posterior survivor function (lifetime).

Finally, we note that forecasts for quantities such as life expectancies, median lifetime, joint (for two people) lifetime and the probability of the first who dies between two people could be obtained easily from the output of the proposed MCMC algorithms as well.

\subsection{Comparisons with existing methods}
\label{comps}

We compare our forecasts of future mortality rates
with forecasts obtained with a series of popular models available in the R package ``StMoMo'' \citep{millossovich2017stmomo}.
The ``StMoMo'' package provides a set of functions for defining and fitting  an abstract model from the family of generalized age-period-cohort stochastic mortality models. For a fitted model the package provides functions for forecasting future mortality rates. In order to quantify the uncertainty of the projections arising from the estimation of the parameters of a model, the package provides also functions for the implementation of bootstrap (semiparametric or on residuals) techniques 
as it was suggested by \cite{brouhns2005bootstrapping}, \cite{koissi2006evaluating} and \cite{renshaw2008simulation}.
Here, we compare predictions for mortality rates obtained using our Bayesian methods with predictions obtained using three commonly used stochastic mortality models.
These are the Lee-Carter (LC) model \citep{lee1992modeling} presented by equation \eqref{eq:lee}, the age-period-cohort model (APC) defined by equation \eqref{eq:apc} and the model of \cite{plat2009stochastic} (PLAT) which combines the model of \cite{cairns2006two} presented by equation \eqref{eq:cairns} with some features of the LC model. 

To perform a full-fledged quantitative evaluation of the forecasts obtained using the different models we used mortality data from UK-Wales and from New Zealand. The New Zealand dataset was included because of the well known \citep{li2014quantitative} characteristic of mortality studies that data from a small country are more comparable with data of insurance portfolios and pension plans. New Zealand had a population of $4.4$ millions on $2011$, quite smaller than the corresponding population of UK-Wales at the same year which was $56.1$ millions.     

The procedure that we used to compare the predictive performance of our proposed models with the performance of the competitive models proceeds as follows. First we obtained from the \cite{HMD} the number of women being 
at risk and the corresponding number of deaths both for UK-Wales and New Zealand during the years $1980-2013$. We used the formula $n_{zt} \approx N_{zt} + \frac{1}{2}d_{zt}$ to transform the average, over the $t$th year, number of people at risk $N_{zt}$ to the initial exposed to risk $n_{zt}$.
Then, for a fixed prediction horizon of $k=5$ and $k=15$ years ahead and for each year $T=1989,\ldots,2013-k$ we used training data of $10$ years, from year $T-9$ up to year $T$, to predict the death probabilities of women with age $z=0,\ldots,89$ years old at the year $T+k$. With the described procedure we obtained, for each of the models, $25-k$ forecasts in the form of prediction intervals each of them at a prediction horizon of $k$ years ahead. 
Based on the conclusions of \cite{currie2016fitting} we we used the logit link for the death probabilities in order to fit the LC, APC and PLAT models.
The details from the implementation of the MCMC algorithms that we used to obtain predictions with the proposed models are given in the on-line supplementary material.

To assess the quality of the obtained prediction intervals for the future death probabilities we calculated the empirical coverage probabilities of the obtained prediction intervals, the mean width of the prediction intervals and the mean interval score. The quality of the mean forecasts was assessed using the root mean squared error of the predicted means. 
For a fixed prediction horizon $k$ and age $z$ the empirical coverage probability of the prediction interval obtained from a given model has been computed as the proportion of the $25-k$ intervals that include the observed probability of death at age $z$ at the year $T+k$, for $T=1989,\ldots,2013-k$. The mean width of the prediction interval is the sample mean of the $25-k$ widths of the obtained prediction intervals and the mean interval score is the sample mean of the scoring rule called interval score; see equation $(43)$ in \cite{gneiting2007strictly}. As it is explained in \cite{gneiting2007strictly} the interval score is a scoring rule which rewards the forecaster who obtains narrow prediction intervals and incurs a penalty, proportional to the significance level of the interval, if the observation misses the prediction interval. This means that we would like to obtain prediction intervals with low mean interval score. See also the on-line supplementary material of the present paper for a more detailed presentation of the interval score.    

\begin{figure}
 \centering
\includegraphics[scale=0.55]{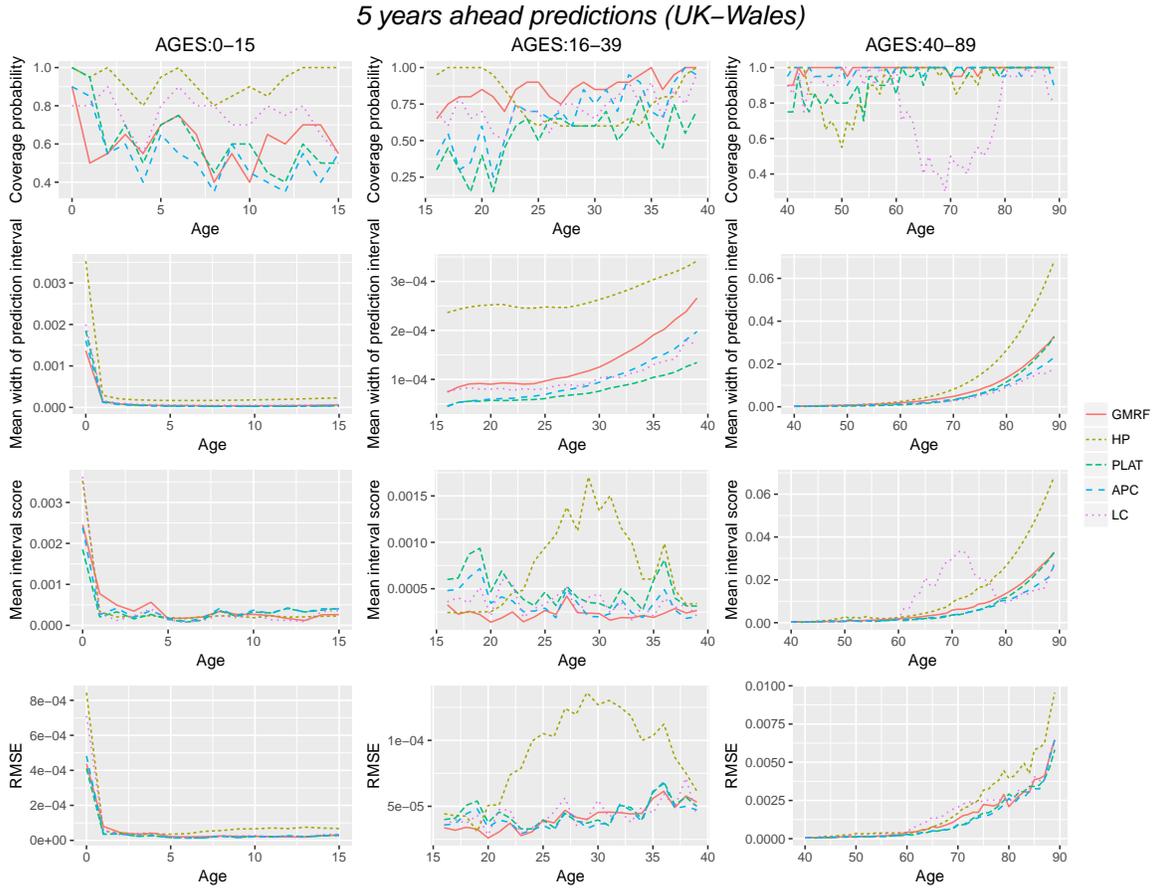}
\caption{Empirical coverage probabilities (first row), mean widths (second row) and mean interval scores (third row) of the $95\%$ prediction intervals and root mean squared error (RMSE) of the mean forecasts (fourth row) calculated for each of the models under comparison by using training UK-Wales mortality data of females from the year $T-9$ until the year $T$, for each $T=1989,\ldots,2008$, to predict the death probabilities of the year $T+5$ across the ages $0-89$ years old. Each column refers to a different age group.}
	\label{UK5}
\end{figure}

\begin{figure}
 \centering
\includegraphics[scale=0.55]{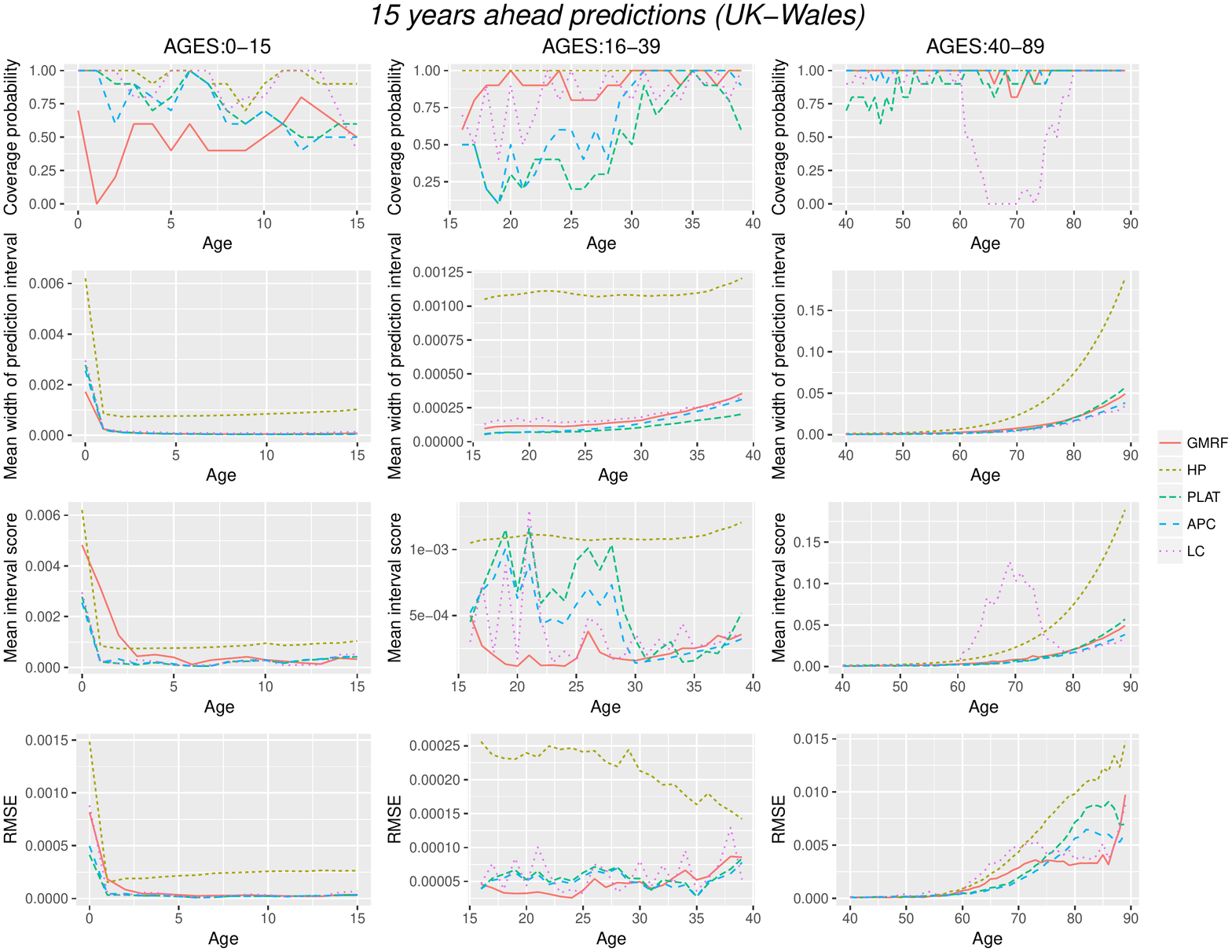}
\caption{Empirical coverage probabilities (first row), mean widths (second row) and mean interval scores (third row) of the $95\%$ prediction intervals and root mean squared error (RMSE) of the mean forecasts (fourth row) calculated for each of the models under comparison by using training UK-Wales mortality data of females from the year $T-9$ until the year $T$, for each $T=1989,\ldots,1998$, to predict the death probabilities of the year $T+15$ across the ages $0-89$ years old. Each column refers to a different age group.}
	\label{UK15}
\end{figure}

\begin{figure}
 \centering
\includegraphics[scale=0.55]{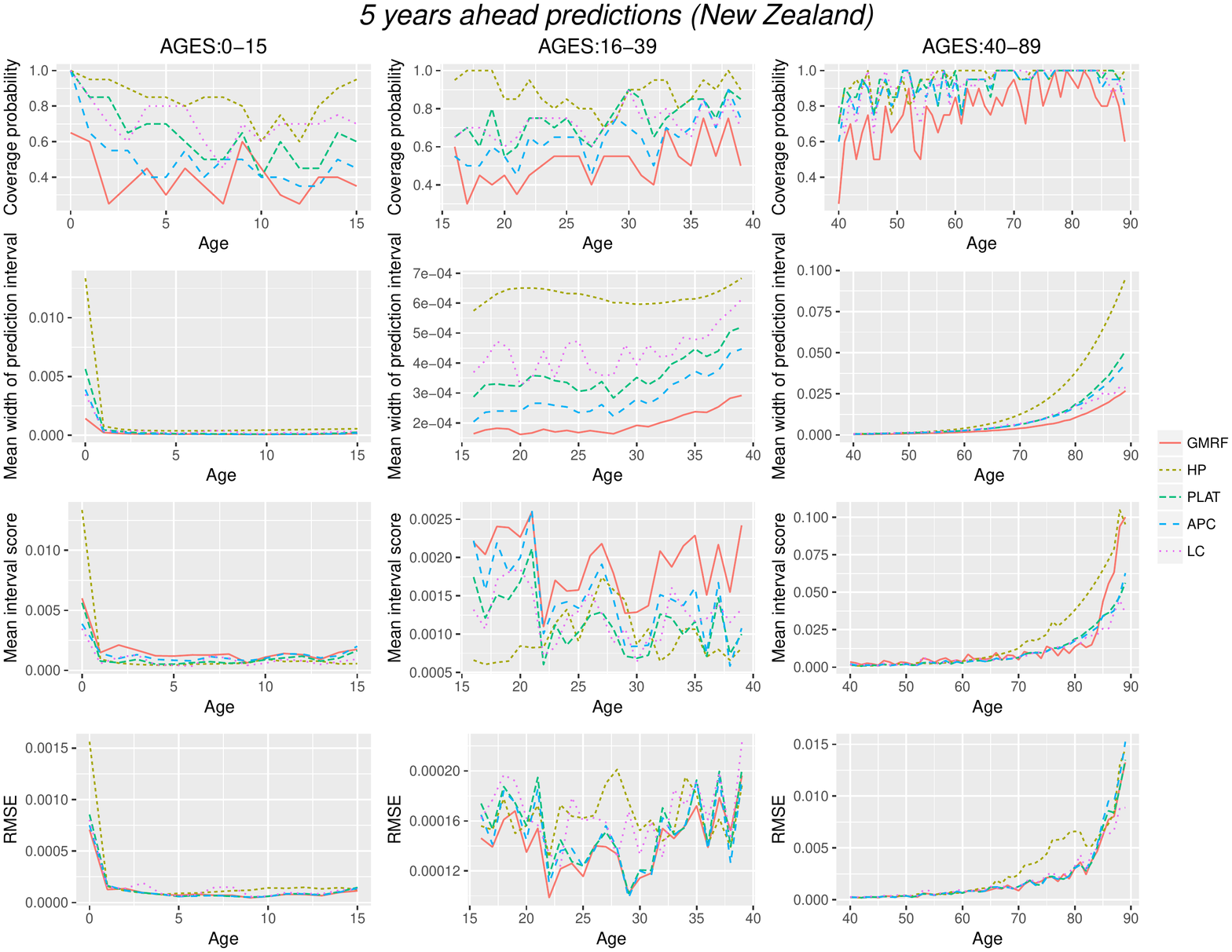}
\caption{Empirical coverage probabilities (first row), mean widths (second row) and mean interval scores (third row) of the $95\%$ prediction intervals and root mean squared error (RMSE) of the mean forecasts (fourth row) calculated for each of the models under comparison by using training New Zealand mortality data of females from the year $T-9$ until the year $T$, for each $T=1989,\ldots,2008$, to predict the death probabilities of the year $T+5$ across the ages $0-89$ years old. Each column refers to a different age group.}
	\label{NZ5}
\end{figure}

\begin{figure}
 \centering\includegraphics[scale=0.55]{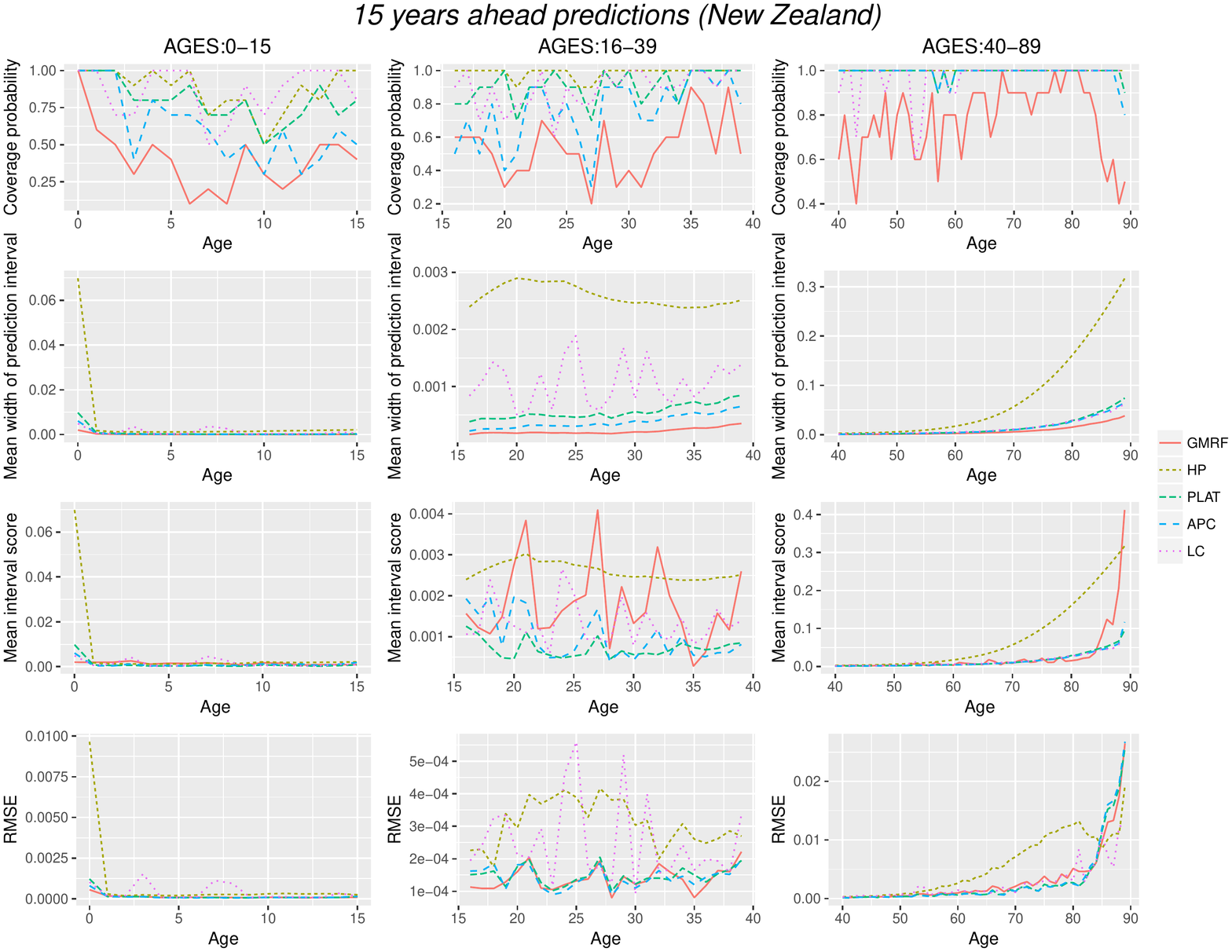}
\caption{Empirical coverage probabilities (first row), mean widths (second row) and mean interval scores (third row) of the $95\%$ prediction intervals and root mean squared error (RMSE) of the mean forecasts (fourth row) calculated for each of the models under comparison by using training New Zealand mortality data of females from the year $T-9$ until the year $T$, for each $T=1989,\ldots,1998$, to predict the death probabilities of the year $T+15$ across the ages $0-89$ years old. Each column refers to a different age group.}
	\label{NZ15}
\end{figure}

Figures \ref{UK5} and \ref{UK15} visualize the evaluation of the $95\%$ prediction intervals obtained from the models under comparison for the UK-Wales dataset. It seems that for the majority of the ages in the range $10-50$ years old the proposed non-isotropic GMRF model delivers the most satisfactory predictions, both for prediction horizons of $5$ and $15$ years ahead, while for ages after $60$ the APC and PLAT models exhibit slightly better predictive performance. Figures \ref{NZ5} and \ref{NZ15} depict the evaluation of the predictions obtained by the models under comparison in the case of mortality data from New Zealand. For an horizon of $5$ years ahead the predictions of the Heligman-Pollard model are more accurate than those obtained from the LC, the APC and the PLAT models for most of the ages up to $60$ years old, while for predictions of $15$ years ahead the APC and PLAT models exhibit the best predictive performance for almost the whole age range. 

In Table \ref{tab01} we summarise the results presented in Figures \ref{UK5}-\ref{NZ15} by providing averages, over ages, of the four measures that we used to assess the predictions obtained from the models under comparison. The proposed non-isotropic GMRF model dominates the Heligman-Pollard model in all the measures that we used except that from the coverage probabilities in the case of the New Zealand dataset.  Nevertheless, even in this case the superiority of the Heligman-Pollard model is quite unimportant since it is based on very wide prediction intervals which have little practical importance.  Moreover, Bayesian inference for the parameters of the dynamic Heligman-Pollard model requires a lot of prior information while inference for the GMRF model is feasible with non-informative priors.  In summary, we propose the use of the GMRF model except if one wishes to relax the stationarity assumptions of the evolution of the mortality curves over the years via the Heligman-Pollard model.

Table \ref{tab01} indicates that the GMRF model, the APC and the PLAT models deliver similar and the most reliable predictions of future death probabilities. Our developed algorithms are not computationally expensive and this is in contrast with existing Bayesian methods for which \cite{li2014quantitative} notes that can take up to a couple of days to run.
Thus, they have the usual advantages of Bayesian inference paradigm,  the most relevant of which is that they can easily be used for projecting, via predictive density functions, of survival probabilities, life expectancies and several other quantities of interest to actuaries and demographers.  Moreover, they can be used routinely in cases with missing data (incomplete life-tables) as it has been demonstrated in \cite{dellaportas2001bayesian} by simply inputing the missing data conditional on the parameters and then, conditional on the missing data, proceeding as described in this article.  With respect to the MCMC mixing behaviour, the imputation of the missing data in the dynamic settings of this paper may be a bit tricky, and may vary between our two proposed models,  since their full conditional density depends not only on the aggregated mortality rates of that year but also on the possibly unobserved mortality rates at the same age of other years.

\section{Conclusions}

We have proposed two models for  forecasting mortality rates.  We have first taken up the theme in \citet{dellaportas2001bayesian} that there are a few attempts at modelling the time evolution of the Heligman-Pollard formula and we proposed a model that does not respect stationarity in the dynamic modelling of the parameters. We have also proposed a non-parametric model based on non-isotropic Gaussian Markov random fields. The evaluation of the forecasts obtained from the proposed and from existing models provides evidence that there are advantages in predicting future mortality rates using our Bayesian models.

Finally we note that there is an increasing interest in the literature for the joint modelling of two or more populations; see, for example, \cite{cairns2011bayesian} and \cite{de2016coherent}. Both our proposed models can be extended towards this direction by modelling the dependence of the different populations using the latent Gaussian processes  of the proposed models. In the case of the Heligman-Pollard model we can assume that the Gaussian density in \eqref{eq:truncrw} is a $16$-dimensional density with the covariance matrix $\bm{\Sigma}$ capturing dependencies of the parameters of the two populations. In the case of the GRMF model the dimension of each $\x_t$ could be a $(2\omega+2)$-dimensional vector resulting in a $(2\omega+2)T \times (2\omega+2)T$ precision matrix in \eqref{eq:matrix} which could be modelled by constructing a non-isotropic GMRF of higher order, see for example in Chapter $3$ of \cite{GMRFbook}.

It is well known in the demographic literature, see for example \cite{renshaw2008simulation},  that it is quite important for demographers, insurance companies and pension institutes the uncertainty of the projections of future mortality rates to be quantified through the computation of prediction intervals. Our proposed Bayesian methodology clearly addresses this issue by producing predictive densities of future data. This feature, together with the fact that one can produce any predictive quantities of interest with simple manipulations of our MCMC output, make our predictions very valuable to actuaries and demographers alike. 

\begin{table}
\caption{\label{tab01} Predictive performance of different models: the average predictive measure over the ages $0-89$ is reported.}
\scalebox{0.75}{
 \fbox{%
\begin{tabular}{l||l|c||c|c|c}
&Model&\multicolumn{2}{c|}{UK-Wales}&\multicolumn{2}{c}{New Zealand}\\
\cline{3-6}
& &$5$ years ahead&$15$ years ahead&$5$ years ahead&$15$ years ahead\\
\hline
\multirow{5}{*}{\parbox{5cm}{Empirical coverage \\probability of prediction intervals}}&GMRF&$0.89$&$0.88$&$0.64$&$0.65$\\
&Heligman Pollard&$0.87$&$0.99$&$0.92$&$0.98$\\
&Lee Carter&$0.76$&$0.77$&$0.84$&$0.93$\\
&APC &$0.82$&$0.86$&$0.77$ &$0.86$ \\
&PLAT &$0.76$&$0.78$&$0.83$ &$0.94$ \\
\hline
\multirow{5}{*}{\parbox{5cm}{Mean width of \\prediction intervals}}&GMRF&$0.004$&$0.006$&$0.003$&$0.005$\\
&Heligman Pollard&$0.007$&$0.019$&$0.010$&$0.022$\\
&Lee Carter&$0.003$&$0.004$&$0.005$&$0.009$\\
&APC &$0.003$&$0.005$&$0.005$ &$0.008$ \\
&PLAT &$0.003$&$0.006$&$0.006$ &$0.009$ \\
\hline
\multirow{5}{*}{\parbox{5cm}{Mean interval score of \\prediction intervals}}&GMRF&$0.004$&$0.006$&$0.008$&$0.012$\\
&Heligman Pollard&$0.008$&$0.019$&$0.011$&$0.022$\\
&Lee Carter&$0.006$&$0.016$&$0.006$&$0.009$\\
&APC &$0.003$&$0.005$&$0.006$&$0.009$ \\
&PLAT &$0.004$&$0.006$&$0.006$ &$0.009$ \\
\hline
\multirow{5}{*}{\parbox{5cm}{Root mean \\squared error }}&GMRF&$0.0008$&$0.0011$&$0.0012$&$0.0018$\\
&Heligman Pollard&$0.0010$&$0.0023$&$0.0015$&$0.0024$\\
&Lee Carter&$0.0009$&$0.0015$&$0.0012$&$0.0016$\\
&APC &$0.0007$&$0.0012$&$0.0013$ &$0.0017$ \\
&PLAT &$0.0007$&$0.0015$&$0.0012$ &$0.0017$ \\
\end{tabular}
}}
\end{table}

\section*{Acknowledgements}
We are grateful to the reviewers for valuable suggestions on presentation and modelling and to Michalis Titsias for many helpful discussions.
This research has been co-financed by the European Union (European Social Fund - ESF)  
and Greek national funds through the Operational  Program ``Education and Lifelong Learning'' of the National Strategic Reference Framework (NSRF) - Research Funding Program: ARISTEIA-LIKEJUMPS-436, and by the Alan Turing Institute under the EPSRC grant EP/N510129/1.

\bibliographystyle{Chicago}

\bibliography{refs}

\end{document}